\documentstyle[11pt,epsfig]{article}
[1999/03/29 PSNFSS v.7.2
Times font as default roman
: S Rahtz]

\begin{document}
\renewcommand{\thefootnote}{\fnsymbol{footnote}}
\sloppy
\newcommand{\rp}{\right)}
\newcommand{\lp}{\left(}
\newcommand \be  {\begin{equation}}
\newcommand \bea {\begin{eqnarray}}
\newcommand \ee  {\end{equation}}
\newcommand \eea {\end{eqnarray}}

\title{Remarks on reply to Johansen's comment :-)}
\author{Anders Johansen
\\
The Niels Bohr Institute, University of Copenhagen\\
Blegdamsvej 17, DK-2100 Kbh. \O, Denmark \\
e-mail: johansen@nbi.dk \hspace{5mm}  URL: http//:www.nbi.dk/\~\/johansen
}
\maketitle
\thispagestyle{empty}
\pagenumbering{arabic}

Any reader of the paper by Laloux {\it et al.} \cite{Laloux} would have been left utterly 
confused after consulting the time series for the price of Japanese Government Bonds (JGB) 
around May 1995 looking for a log-periodic power law acceleration in the data. The authors 
of \cite{Laloux} now confess their error and admit that the prediction date given in \cite{Laloux} 
was incorrect and that the correct prediction was for August 1995. However, they now report 
\cite{reply} that the analysis leading to the prediction was made in May 1995. Obviously, this 
is {\it also} incorrect since the data used in the analysis, see fig. 1 of \cite{comment}, does
not start before June 1995. That the authors of the reply (LMAB) have severe problems regarding 
dating certain events becomes even clearer reading the second section of the reply. Reference 
[3] of the reply (corresponding to reference \cite{SJ2001}) is essentially a 19 page answer to 
the criticism put forward in \cite{Feig}. Nevertheless, LMAB writes that I and Didier Sornette 
(JS) ignores the work of J.A. Feigenbaum!

Let us now turn away from the prediction experiment in question as well as the issue of
whom cites whom and address the physical issues raised by LMAB. That crashes occurs for
a number of different reasons and unfolds in various ways is quite obvious. For example,
the bond crash of October 1998 mentioned by LMAB is generally explained as a result of
the heavy investments made by German Banks in Russia. When the Rubel crashed, so did
the Bundes bond. Other crashes triggered by external events are the ones on Wall Street
in 1973 (OPEC oil embargo) and 1974 (Nixon's resignation). However, no consensus regarding
the origin of the world-wide crash of 1987 exists. This clearly illustrates that crashes
occurs for a variety of reasons of which speculative bubbles are just one. What we (JS) has
argued in a number of papers \cite{SJ2001,outl} is that speculative bubbles on the financial
markets are more often than not quantifiable by a log-periodic power law acceleration ending 
in a crash or a large correction. That the crash/correction is not certain was already made 
clear in 1998 \cite{JLS} where a probabilistic framework was developed. I wish to stress 
that such a probabilistic framework is {\it essential} in order for our hypothesis to make
sense. In the same paper, an explanation to why that the time of the crash/correction predicted 
by the governing log-periodic power law  eq. {\it in general} over-shoots the actual date was 
also offered.

With respect to the work on the distribution of price changes and the possible existence of 
outliers, my own work with D. Sornette has been done using {\it drawdowns} on daily data whereas 
the work by V. Plerou {\it et al.} and J.-F. Muzy {\it et al.} mentioned by LMAB was done using 
returns on time series containing intra-day data. A comparison of results is therefore very 
difficult, since returns are calculated over a {\it fixed} time horizon whereas drawdowns uses
a {\it flexible} time horizon adapted to the market dynamics \cite{outl}. What D. Sornette and I 
have argued is that drawdowns is a more natural and relevant measure of {\it e.g.} stock market 
fluctuations than returns on an arbitrary fixed time scale.

Last, a few minor points should be addressed. First, using the same statistics for the drawups
as for the drawdown the largest drawup shown in fig. 1 of \cite{comment} is $4.2\%$. Hence, there
are no ``drawup outlier'' in the data shown. Second, the remark of LMAB that ``put options are 
worth nothing if the contract is above the exercise price'' is only true on the maturity date 
itself. Considering the nature of option trading it is quite extraordinary that no money was lost.
In fact, this was one of the reasons why the experiment was considered a partial success at that
time \cite{DS}. This was certainly due to the fact that the price was declining unusually fast even 
though the JGB did not crash in a conventionally sense. Last I consider it quite strange that one
of the authors of \cite{Laloux} (R. Cont) is silent about my comment.


\begin{thebibliography}{}
\bibitem{comment} Comment on ``Are financial crashes predictable''.
\end{thebibliography}
\end{document}